\documentclass[runningheads]{llncs}
\usepackage{mathtools, amssymb, amsfonts}

\usepackage{nicefrac}

\usepackage[printonlyused]{acronym}
\usepackage{graphicx}

\usepackage{listings}
\usepackage{xcolor}
\lstdefinelanguage{turtle}{
 keywords={@prefix, a, rdf, rdfs, owl},
 keywordstyle=\color{blue},
 basicstyle=\ttfamily\footnotesize,
 morecomment=[l][\color{gray}\ttfamily]{\#},
 morestring=[b][\color{red}]{\"},
}
\usepackage[hang,small,bf,tight]{subfigure}
\usepackage{url} 

\usepackage{tikz}
\usetikzlibrary{shapes.geometric, arrows, positioning}
\usepackage{array}
\usepackage{booktabs}
\usepackage{multirow}
\usepackage{colortbl}
\usepackage{spreadtab}
\usepackage{longtable}
\usepackage{pdflscape}
\usepackage{float}
\usepackage[super]{nth}
\usepackage{enumitem}
\usepackage{todonotes}
\usepackage{footmisc}
\usepackage{minted}
\usepackage{indentfirst}
\usepackage{cite}
\usepackage{ulem}
\usepackage[T1]{fontenc}
\usepackage[utf8]{inputenc}
\usepackage[english]{babel}
\usepackage{iflang}
\addto\captionsenglish{
 \renewcommand{\contentsname}{\null ~\vskip 1.em \hfil Contents\hfil}
 \renewcommand{\refname}{Bibliography}
}
\usepackage{graphicx}
\setlist[itemize]{label=\textbullet}

\begin{document}
\title{Ontology-Driven Model-to-Model Transformation of Workflow Specifications}
\titlerunning{Ontology-Driven Transformation}
%
\author{Francisco Abreu\inst{1}\orcidID{0009-0007-3458-2677} \and
Luís Cruz\inst{1}\orcidID{0009-0002-2832-9791} \and
Sérgio Guerreiro\inst{1,2}\orcidID{0000-0002-8627-3338}}
\authorrunning{F. Abreu et al.}
%
\institute{Instituto Superior Técnico, Universidade de Lisboa,\\
Av. Rovisco Pais 1, 1049-001 Lisboa, Portugal
\and
INESC-ID, Lisbon, Portugal\\
\email{\{francisco.abreu,luis.cruz,sergio.guerreiro\}@tecnico.ulisboa.pt}}

%
\maketitle 
\begin{abstract}
Proprietary workflow modeling languages such as Smart Forms \& Smart Flow hamper interoperability and reuse because they lock process knowledge into closed formats. To address this vendor lock-in and ease migration to open standards, we introduce an ontology-driven model-to-model pipeline that systematically translates domain-specific workflow definitions to Business Process Model and Notation (BPMN) 2.0. The pipeline comprises three phases: RML-based semantic lifting of JSON to RDF/OWL, ontology alignment and reasoning, and BPMN generation via the Camunda Model API. By externalizing mapping knowledge into ontologies and declarative rules rather than code, the approach supports reusability across vendor-specific formats and preserves semantic traceability between source definitions and target BPMN models. We instantiated the pipeline for Instituto Superior Técnico (IST)’s Smart Forms \& Smart Flow and implemented a converter that produces standard-compliant BPMN diagrams. Evaluation on a corpus of 69 real-world workflows produced 92 BPMN diagrams with a 94.2~\% success rate. Failures (5.81~\%) stemmed from dynamic behaviors and time-based transitions not explicit in the static JSON. Interviews with support and development teams indicated that the resulting diagrams provide a top-down view that improves comprehension, diagnosis and onboarding by exposing implicit control flow and linking tasks and forms back to their sources. The pipeline is generalizable to other proprietary workflow languages by adapting the ontology and mappings, enabling interoperability and reducing vendor dependency while supporting continuous integration and long-term maintainability. The presented case study demonstrates that ontology-driven M2M transformation can systematically bridge domain-specific workflows and standard notations, offering quantifiable performance and qualitative benefits for stakeholders.

\keywords{Ontology‑driven model transformation \and Business Process Model
 and Notation (BPMN) \and Semantic lifting \and RDF Mapping Language (RML) \and
 Model‑driven engineering \and Workflow translation \and Smart Forms \& Smart Flow}
\end{abstract}

\section{Introduction}\label{sec:introducion}

The integration of business process modeling and semantic technologies is increasingly important in enterprise information systems where automation, interoperability and knowledge extraction are critical for organizational efficiency. Although our translation targets Business Process Model and Notation (BPMN) 2.0, the approach is designed to be adaptable to other workflow schemas. BPMN 2.0 is a widely adopted standard that provides a graphical notation for business processes together with execution semantics~\cite{BPMN_2.0.2}. It facilitates communication between business and technical stakeholders and promotes tool interoperability through a well‑defined meta‑model and extension mechanism~\cite{BPMN_2.0.2, Fundamental_of_BPMN_Dumas_2018}. Its extensibility allows proprietary information to be encapsulated in vendor‑specific namespaces while preserving the core structure~\cite{BPMN_2.0.2}.

Many organizations nevertheless rely on proprietary workflow notations that lock process knowledge into closed formats, hindering integration, analysis and reuse. At Instituto Superior Técnico (IST), the Smart Forms \& Smart Flow modules are used to digitize institutional processes. Smart Forms allow users to submit online forms and obtain digital documents. Form definitions and submissions are specified in JSON, and outputs can be delivered electronically or collected at service desks. Smart Flow is a workflow engine based on queues and templates specified in JSON, coupled with Java processors. It provides interfaces for defining workflows, visualizing them and executing processes. The coexistence of form‑centred and activity‑driven paradigms leads to semantic mismatches and technical barriers, often forcing duplication and inconsistency. These modules are built on top of the open source \texttt{fenix} framework\footnote{\url{https://fenixedu.org/dev/bennu/fenix-framework/fenix-framework}}, but the workflow specifications themselves remain closed. Even when they eventually become publicly available, the challenge of interoperating with other tools persists.

Proprietary “executable BPMN” dialects illustrate that vendor lock-in is not unique to the IST context. Camunda and Oracle embed engine‑specific data mappings and control information in extension elements, overriding the base BPMN specification and making models executable only on their own engines~\cite{cib_seven_online, BPMN_2.0.2, Fundamental_of_BPMN_Dumas_2018}. Oracle even distributes pre‑built virtual machines for its SOA/BPM stack~\cite{oracle_prebuilt_vm}, and its products are licensed at a higher cost than open-source tooling. Differences between versions of the same vendor (e.g., Camunda 7 vs.\ Camunda 8)\footnote{\url{https://docs.camunda.io/docs/guides/migrating-from-camunda-7}} necessitate migration and re‑work~\cite{concept_diff_camunda_8_online, Fundamental_of_BPMN_Dumas_2018}. Cross‑vendor migrations (such as from Oracle SOA to Camunda) magnify the effort~\cite{semantically_2012_abramowicz}, and evolving BPMN profiles and vendor extensions can trigger repeated translation of legacy models~\cite{bpmna_online}. These realities underscore that execution concerns are engine‑specific and highlight the need for approaches that localize variability and minimize vendor dependency\footnote{\url{https://www.oracle.com/cloud/price-list/}}. This shows that BPMN permits vendor‑specific extensions - tool vendors insert proprietary information into BPMN files via extension elements, and this information is not guaranteed to be interpreted uniformly by other tools~\cite{bpmn_interchange_kurz_2016}.
The product dependency is clear: Oracle’s SOA Suite behaves similarly to the Smart Form \& Smart Flow environment in this regard. The solution developed \textbf{aims to be generalizable} to such cases.

We address these issues by introducing an ontology‑driven model‑to‑model (M2M) pipeline that translates proprietary workflow definitions into standard BPMN 2.0 diagrams. An ontology‑driven mapping localizes variability in declarative rules instead of scattered code, and the translation of JSON definitions into BPMN provides stakeholders with a standard, visual representation they can understand, unlike raw code or ad‑hoc simulations. Our pipeline comprises three phases: (1) \emph{semantic lifting} of JSON specifications into RDF using the RDF Mapping Language (RML), (2) \emph{ontology integration and reasoning} to align the lifted data with a BPMN ontology, and (3) \emph{model generation}, which produces BPMN diagrams via the Camunda Model API\footnote{\url{https://javadoc.io/doc/org.camunda.bpm/camunda-engine/7.22.0/index.html}}. The implementation is used as a case study of a general transformation scheme and demonstrates how semantics and traceability can be preserved throughout the process.

\subsection{Contributions}\label{sec:contributions}

The ontology‑driven pipeline supports alignment of models from diverse systems through a shared vocabulary, enabling semantic queries and cross‑platform compatibility. By adapting the ontology and mapping rules, it can be extended to workflow languages beyond Smart Forms \& Smart Flow. Although ontology‑driven adoption requires an initial investment in mapping and ontology design, it offers long‑term benefits such as enhanced interoperability and reusability~\cite{hamri2014semantic, yaml_aalst_2005}.
\begin{itemize}
 \item We formulate a general ontology‑driven pipeline for translating proprietary workflow definitions to BPMN. The pipeline comprises semantic lifting with RML, ontology integration and reasoning, and programmatic model generation. The design emphasizes reusability and generalization beyond the studied implementation.
 \item We instantiate the pipeline for Smart Forms \& Smart Flow, providing an operational converter that produces BPMN diagrams via the Camunda Model API. The implementation serves as a representative case study that illustrates the phases and boundary conditions of the approach.
 \item We evaluate the pipeline on a corpus of 69 workflow specifications from IST, producing 92 BPMN diagrams. The evaluation reports quantitative performance and qualitative insights. The results show high translation accuracy (94.2~\% successful conversions) and sub‑second performance, with an average translation time of 404~ms per file.
 \item We discuss generalization, threats to validity and opportunities for future research, situating our work within the broader literature on M2M transformations and semantic workflow engineering. The empirical indications of improved maintainability, scalability and semantic traceability are framed as projectable under stated boundary conditions~\cite{what_is_an_ontology_2009,ontology_smith_2012}.
\end{itemize}

\subsection{Significance \& Research Gap}\label{sec:significance}
It is important to articulate why this contribution is both significant and timely for the Software and Systems Modeling community. In practice, many organizations rely on proprietary workflow engines that lock process knowledge into closed formats, forcing re‑implementation or manual reverse engineering when moving to open standards, as mentioned before. Existing model‑to‑model translators often hard‑code mappings in imperative code, offer limited semantic alignment and traceability, or focus on a single vendor without addressing generalization. Our ontology‑driven pipeline fills this gap by externalizing mapping knowledge into ontologies and declarative RML rules, aligning lifted data with a BPMN ontology through reasoning, and generating standard compliant diagrams with automatic layout. This combination of semantic lifting, ontology integration, reasoning and programmatic generation supports interoperability across vendors and preserves traceability from source JSON to target BPMN. It therefore provides a replicable path toward vendor‑independent process models and demonstrates that ontology‑driven techniques can bridge closed workflow languages to open standards.

The remainder of this paper is structured as follows. Section~\ref{sec:background} summarizes background on BPMN, semantic lifting and ontologies. Section~\ref{sec:related} reviews related work on model‑to‑model transformations and workflow translation. Section~\ref{sec:method} describes our ontology‑driven transformation method, including its design principles and implementation. Section~\ref{sec:evaluation} presents the evaluation results. Section~\ref{sec:discussion} reflects on threats to validity and relates our contributions to the existing literature. Section~\ref{sec:conclusion} concludes and outlines future work.

\section{Background}
\label{sec:background}

BPMN 2.0 is an Object Management Group (OMG) standard that defines a graphical notation for business processes together with execution semantics~\cite{BPMN_2.0.2}. It distinguishes flow objects (events, activities, gateways), connecting objects (sequence flows, message flows), swimlanes and artifacts. The standard seeks to provide a common language between business and technical stakeholders and to serve as a bridge to executable process formats~\cite{Fundamental_of_BPMN_Dumas_2018, Ontology_Analysis_BPMN_Singer_2019}. Its meta‑model and execution semantics are formally defined in the specification, and its standardized serialization enables the exchange of process definitions across vendor tools~\cite{BPMN_2.0.2}. BPMN’s expressiveness and widespread tool support have made it the de facto standard for workflow modeling. However, when tools incorporate proprietary extensions, they depart from the core semantics and compromise portability.~\cite{BPMN_2.0.2}.

Smart Forms \& Smart Flow are composed of JSON definitions and Java processors. They are distributed across loosely coupled files, increasing modeling complexity and obscuring an end‑to‑end view of the process. The challenge lies in the complexity of the system, which continues to evolve. Model‑to‑model transformations aim to address such challenges by translating source into target models within a model‑driven engineering context. Ontology‑driven transformations are less explored; aligning models via ontologies can provide a semantic bridge between heterogeneous languages.

\paragraph*{Semantic Lifting \& ontologies.}
Semantic lifting converts structured data into an explicit knowledge representation such as RDF or OWL. The RDF Mapping Language (RML) is a declarative language that maps heterogeneous sources (e.g., JSON or CSV) to RDF triples~\cite{rml_mapper_java_2016}. RML mappings specify how subjects and predicate - object pairs are generated using iterators and mapping rules. Applying RML to the Smart Forms \& Smart Flow definitions yields an RDF graph aligned with domain ontologies\footnote{\url{https://rml.io/specs/rml/}}. Ontologies provide shared vocabularies and formal semantics, enabling reasoning over the lifted data. The Web Ontology Language (OWL) supports formal modeling and alignment through constructs such as \texttt{owl:equivalentClass}~\cite{equiClassProperty2013}. 

Our approach builds upon a BPMN ontology that extends previous semantic representations of BPMN, notably the Business Process Ontology (BBO)\footnote{\url{https://www.irit.fr/recherches/MELODI/ontologies/BBO}}. This ontology serves as the target schema for the lifted models~\cite[Chapter 10]{Fundamental_of_BPMN_Dumas_2018}. It allows us to produce semantically precise BPMN models and ensures traceability from the original JSON through to the BPMN output~\cite{bbo_paper_2019}.

\paragraph*{Diagram Layout \& visualization.}
The solution includes a diagram visualization component inspired by the approach of Kitzmann et al.~\cite{layout_algorithm}. The original algorithm\footnote{\url{https://github.com/process-analytics/bpmn-layout-generators}} offered basic diagram layout; our improved module adds support for BPMN collaboration features such as pools, participants and swimlanes. The layout adopts a layered, left‑to‑right progression with gateway symmetry, minimal edge crossings and waypoint minimization to enhance readability. This baseline auto‑layout is deterministic and adequate for continuous integration (CI) artifacts; users remain free to perform manual adjustments in BPMN editors when desired.

Three key insights motivate our ontology‑driven transformation: (i) JSON objects and keys can be lifted into an explicit, structured specification of a domain conceptualization~\cite{what_is_an_ontology_2009}; (ii) Smart Forms \& Smart Flow are independent modules and do not share a one‑to‑one mapping to BPMN - separate ontologies are therefore warranted, with explicit correspondences to BPMN constructs; and (iii) existing workflow specification and transformation methods suffer from a lack of standardization, absence of formal semantics and insufficient support for model validation and visualization. Our method seeks to enable robust automation and knowledge extraction for enterprise process management by addressing these limitations.

\section{Related Work}\label{sec:related}

Model‑to‑model transformations underpin model‑driven engineering by converting artifacts across meta‑models. Early surveys distinguish between endogenous and exogenous transformations, batch and incremental execution, and unidirectional and bidirectional mappings~\cite{sendall2003heart,mens2006taxonomy,czarnecki2006survey}. The OMG Query/View/Transformation (QVT) specification codified declarative and imperative M2M languages~\cite{qvt2005}. Frameworks such as the Atlas Transformation Language (ATL) and the Epsilon Transformation Language (ETL) popularized rule‑based approaches combined with imperative helpers~\cite{jouault2006atl,kolovos2008etl}. More recent hybrid platforms integrate declarative rules with imperative extensions and employ graph databases for scalability. For example, YAMTL and the eMoflon family (Neo/IBeX) use incremental pattern matching and bidirectional TGGs to support efficient synchronization~\cite{mkaouar2022benchmark,weidmann2021emoflonneo,emoflonIBeX}. Benchmarks demonstrate that no single tool dominates across domains; performance and usability vary, and advanced features often require expert tuning~\cite{mkaouar2022benchmark}. Formal correctness is rarely addressed, though frameworks such as KBX have shown that verified bidirectional synchronization is possible~\cite{zhao2024kbx}. These developments suggest that incremental and bidirectional features are desirable yet still maturing.

Beyond software models, numerous works translate between business process notations. Early research mapped BPMN to block‑structured languages such as BPEL by restructuring graphs into well‑nested patterns. Semantic mismatches (e.g., error handling and compensation) limited round‑trip fidelity. Other mappings targeted Petri nets, YAWL and EPCs to enable analysis, execution or migration~\cite{semantic_analysis_dijkman_2008, bpmn_into_yawl_nets_springer_2009}. Despite improved tool support, core challenges remain: differences in control‑flow formalisms, incomplete support for organizational roles and data, and inconsistent treatment of OR‑joins and cancellation regions~\cite{semantic_analysis_dijkman_2008}. Since 2020, contributions have focused on integrating BPMN 2.0 features (event subprocesses, escalation events) and coupling decision models (DMN), rather than proposing fundamentally new transformations~\cite{Fundamental_of_BPMN_Dumas_2018}. This maturity underscores the need for semantic pivots that preserve meaning when translating across notations.

Ontology‑driven approaches address this by grounding models in shared semantics. Common workflow ontologies and the Business Process modeling Ontology (BPMO) annotate BPMN, YAWL and BPEL constructs with concepts such as \emph{Activity}, \emph{Role} and \emph{ControlFlow}, enabling uniform queries and cross‑language interoperability~\cite{hamri2014semantic,cabral2009bpmo}. Recent research lifts BPMN models into OWL 2 to leverage description‑logic reasoning~\cite{kchaou2021bpmn2owl}. Semantic lifting frameworks such as RML map heterogeneous data into RDF, while rule languages (e.g., SWRL, SPARQL) express domain‑specific constraints and infer implicit relationships~\cite{hamri2014semantic}. These techniques support semantic alignment and reasoning at the cost of increased modeling effort and computational overhead.

Our work situates itself at this intersection. We build upon classical M2M languages while prioritizing semantic clarity and traceability. Proprietary Smart Forms \& Smart Flow workflows are declaratively mapped into RDF via RML, aligned to a BPMN ontology through bridge axioms and reasoning, and then programmatically generated into BPMN diagrams. This ontology‑driven pipeline complements template‑based translators by lifting models into a common semantic space and can be extended with incremental or bidirectional mechanisms as they mature. In doing so we aim to bridge the gap between rich semantic approaches and practical workflow translation.

\section{Ontology‑Driven Transformation Method}
\label{sec:method}

\begin{figure}[htpb]
\centering
\includegraphics[width=\textwidth]{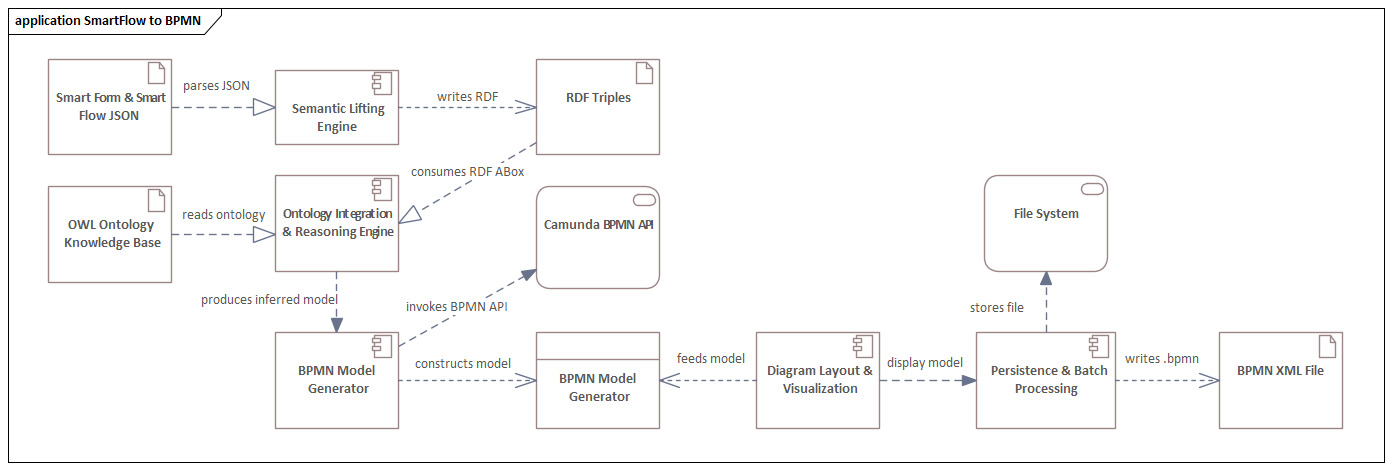}
\caption{Component structure and artifact relationships in the ontology-driven translation pipeline.}
\label{fig:component_model}
\end{figure}

\begin{figure}[htpb]
\centering
\includegraphics[width=\textwidth]{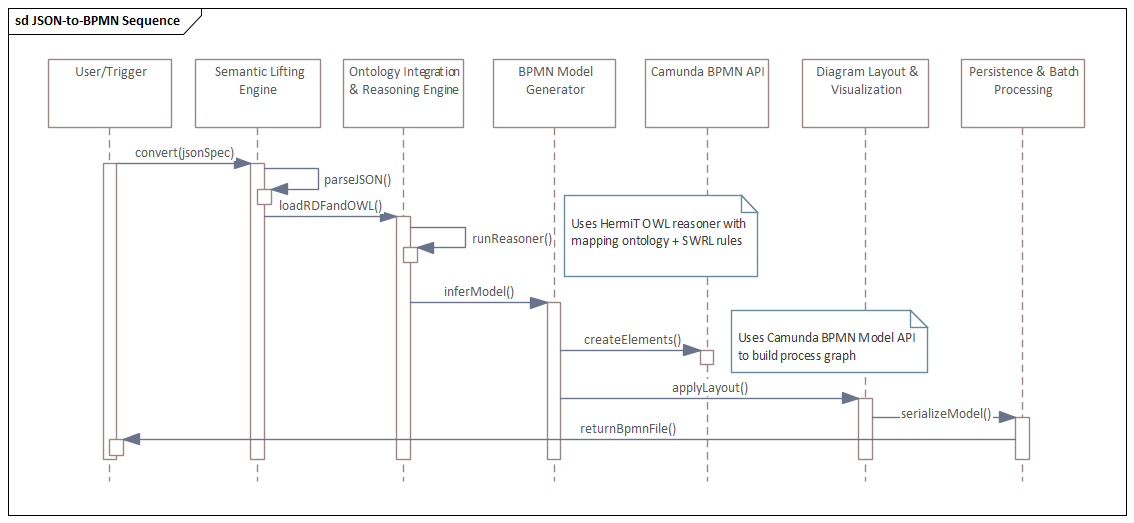}
\caption{Runtime data flow from input selection through lifting, reasoning, generation, layout and persistence.}
\label{fig:sequence_diagram}
\end{figure}

\noindent\textbf{Overview.}
Our pipeline translates Smart Forms \& Smart Flow specifications into BPMN by (i) lifting JSON artifacts to RDF aligned with a domain ontology, (ii) aligning them with a target BPMN ontology via a mapping ontology and reasoning, and (iii) generating a BPMN model with automatic layout and CI-friendly persistence. Figures~\ref{fig:component_model} and \ref{fig:sequence_diagram} depict the component structure and end-to-end flow.

\paragraph{From a code-centric instantiation to an ontology-driven method.}
An initial, code-centric prototype implemented a direct JSON~$\rightarrow$~Java~$\rightarrow$~BPMN pipeline: JSON definitions were parsed into POJOs and lifted into an Intermediate BPMN (IBPM) structure prior to emission with the Camunda BPMN Model API. Although feasible end-to-end, IBPM rapidly accumulated special-case handlers as the specification evolved (e.g., button patterns, conditional targets, multi-instance conventions), requiring code edits rather than configuration and limiting portability beyond a specific engine/version. The ontology-driven approach externalizes mapping knowledge into ontologies and RML rules.

\subsection{Ontology Knowledge Base Construction}
\label{sec:ontology}

We assemble a layered knowledge base comprising three ontologies. A \emph{domain ontology} captures Smart Forms \& Smart Flow concepts (requests, flow templates, action nodes, queues, processors). A \emph{target BPMN ontology}, building on BBO~\cite{bbo_paper_2019,BPMN_2.0.2}, formalizes BPMN 2.0 constructs (processes, tasks, sequence flows). A \emph{mapping ontology} encodes bridge axioms (class/property equivalences and constraints) and, where needed, SWRL rules for non-trivial correspondences. Typical axioms include \texttt{owl:equivalentClass} between \texttt{sf:ActionNode} and a BPMN task type (e.g., \texttt{bpmn:UserTask} when a step is human-executed) and property correspondences linking Smart Flow relations (e.g., \texttt{sf:has\_queue}) to BPMN organizational structures (lanes or pools).

At runtime, the three ontologies and the instance data (ABox) are imported into a single OWL ontology. In description-logic terms, the TBox defines vocabulary (classes/properties) and the ABox contains assertions about individuals; a DL reasoner operates over the TBox to classify ABox individuals and derive implicit correspondences.

\subsection{Semantic Lifting: RML Mapping from JSON to RDF}
\label{sec:lifting}

The lifting step consumes two inputs: (1) Smart Forms \& Smart Flow JSON specifications and (2) the catalogue of Java processors, i.e., named handlers invoked by workflow steps. The JSON definitions are authoritative for nodes, transitions and user‑interface behavior (e.g., buttons), while processors encode system actions or synchronizations that must be retained for traceability. RML mappings specify how JSON objects and arrays become RDF individuals and triples aligned with the domain ontology. Every BPMN element later generated retains a back‑link to its source individual, supporting auditability and explainability.

After lifting, the domain, BPMN and mapping ontologies are merged with the ABox, and a Description‑Logic (DL) reasoner (e.g., HermiT) is applied. The reasoner classifies individuals into BPMN categories (e.g., \texttt{bpmn:UserTask} or \texttt{bpmn:ServiceTask}) and materializes implicit relations (e.g., that multiple guarded transitions imply an exclusive gateway). Optional SWRL rules handle higher‑order patterns such as split/merge gateways \textit{(see actionnode\_generated\_by\allowbreak\_semantic\_lifting.png)}.

We perform semantic and syntactic validation before model generation. Semantic checks - SPARQL and DL queries - ensure (a) knowledge‑base consistency (no unsatisfiable classes), (b) required gateway typing when fan‑out or guard patterns are present, (c) resolvable lane assignment from queues, and (d) presence of traceability IRIs on all elements. Syntactic checks leverage the Camunda BPMN API validators to ensure the model is structurally correct prior to serialization \textit{(see load\_ontologies\_and\_abox.png)}.

\subsection{BPMN Model Generation}
\label{sec:generation}

The generator interrogates the reasoned knowledge base to build an in‑memory BPMN model using the Camunda BPMN Model API. Elements are instantiated only once their semantic pre‑conditions are satisfied: tasks are created from typed individuals; sequence flows arise from \texttt{sf:transitionsTo} relations and inferred gateways; pools and lanes derive from organizational semantics (queues); subprocesses correspond to grouped regions when indicated; and start and end events map to designated entry and exit nodes. Programmatic construction offers precise control over BPMN semantics, immediate access to validators and clean integration with diagram interchange (DI) creation. Because positional and typing decisions have been justified semantically, the generator remains compact and engine‑agnostic \textit{(see generateBpmnModel\_pseudo.java)}.

\subsection{Diagram Layout \& Visualization}
\label{sec:layout}

The generator emits BPMN Diagram Interchange elements (shapes and edges) so that models open with a readable layout in standard editors. We apply a layered, flow‑direction layout that prioritizes: (i) consistent left‑to‑right progression, (ii) gateway symmetry, (iii) minimal edge crossings and (iv) waypoint minimization for readability. This baseline auto‑layout is lightweight, deterministic and adequate for CI artifacts; editors remain free for manual touch‑ups when desired. Lane assignments, gateway placement and conditional labels are materialized visually, making implicit behavior explicit for stakeholders \textit{(see lane\_aware\_auto\_layout.png, comparison\_layout\_kitzmann.png)}. This visibility was a key driver for adoption at IST and was validated in stakeholder sessions.

\subsection{Persistence \& Batch Processing}
\label{sec:persistence}

The pipeline supports headless, directory‑level batch conversion integrated with institutional CI. For each input it produces: (i) a reasoned BPMN XML, (ii) a PNG rendering and (iii) logs containing lifting and validation outcomes and per‑file timings \textit{(see process\_all\_files\_pseudo.java)}. Failed cases are isolated with actionable diagnostics (e.g., missing forms or unsupported scripts). This enables continuous synchronization of BPMN artifacts with evolving JSON specifications and routine distribution of artifacts to support and development teams.

\subsection{Demonstration - Supplementary Material}\label{sec:supplementary-material}

For implementation details beyond the space of this paper, the supplement (demo\allowbreak\_SoSyM.pdf) provides a compact end‑to‑end walkthrough of a representative translation: from Smart Forms \& Smart Flow, through RML lifting to RDF, ontology alignment and reasoning checks, to programmatic BPMN generation with DI layout and traceability. It also includes minimal instructions for running the pipeline and trimmed artifacts (inputs, key mappings and the resulting BPMN) so that readers can replicate the example. The main paper presents the method; operational specifics are deferred to the supplement.

\section{Evaluation}
\label{sec:evaluation}

The evaluation investigates both verification - whether the pipeline was built correctly - and validation - whether the produced BPMN diagrams meet stakeholder needs. All experiments were executed automatically in CI during 1-31 August on a stable cohort of Smart Forms \& Smart Flow source files. From 69 distinct sources, the pipeline produced 92 BPMN diagrams\footnote{The 69 sources include Smart Form (request) and Smart Flow (workflow) definitions; the pipeline outputs one diagram per part. CI logs show 58 request diagrams and 34 flow diagrams (92 total).}. The CI runners used a pinned Docker image (Java 17, Node.js 20), ensuring reproducibility of results \textit{(see evaluation/ folder in zip)}.

\subsection{Verification}\label{subsec:verification}

Verification assessed the structural conformance of the generated BPMN models with the intended semantics, execution performance and resource usage, and robustness on a realistic corpus. Each JSON source was parsed and validated prior to mapping; malformed files were excluded from aggregates and reported. The cohort covered diverse administrative, human‑resources and academic processes from IST’s repository. The CI job builds the translator, converts all workflows, renders BPMN to PNG for inspection and publishes artifacts and logs on each run. Jobs execute on autoscaled runners.

A description‑logic reasoner classifies lifted individuals, checks disjointness constraints and materialises correspondences to BPMN classes (e.g., \texttt{bpmn:UserTask} or \texttt{bpmn:StartEvent}). A per‑process “matrix report” \textit{(matrix\_report.png)} relates Smart Flow elements (\textit{F}) to produced BPMN elements (\textit{B}), exposing omissions or unexpected inflation; across simple to complex flows, counts track predictably with source structure. Behavior‑level conformance in Camunda Modeler remains a manual activity and is future work to automate.

On the full dataset the mean translation time is 404~ms per file (CPU time 406.78 ms), with an average heap delta of 6.82 MB; total batch time for 92 diagrams is 37.2 s. Occasional heap spikes occur in large templates during mapping and reasoning; recommended mitigations include ontology‑service caching and persistent reasoner sessions. Larger JSON sources (over 1,000 lines) complete without timeouts; regression analyses show a clear monotonic trend-runtime grows roughly linearly with model size over the observed range (20-80 elements/ 300-2,600 lines). Automated checks and measurements ensure comparability over time by using stable files and avoiding material schema changes.

The overall success rate across the August corpus is 94.2~\%, with failures concentrated in two patterns: (i) unresolved references to dynamic or undeclared nodes generated at runtime by processors and (ii) time‑based or dynamically activated transitions that are not explicit in the static JSON. These cases yield incomplete static reachability and prevent a complete BPMN. Otherwise the pipeline consistently produces structurally sound BPMN from heterogeneous JSON definitions at sub‑second latencies, scales within the observed complexity envelope, and exhibits a high success rate. Remaining failures stem from runtime‑only behaviors not encoded in the static artifacts and are thus boundary‑condition cases rather than defects of the transformation.

\subsection{Validation}\label{subsec:validation}

\paragraph*{Aim \& design.}
Validation asked whether the translation pipeline meets its intended use. While verification asked whether the pipeline was built correctly, validation asks whether the produced BPMN diagrams are useful for stakeholders. The evaluation was therefore integrated into the operations of IST’s support team (ALU) and development team (DASI). Because ALU acts as the intake point for user requests and escalates to the back office when needed, the evaluation targeted both front-line triage and in-depth diagnostic tasks. The focus was on assessing usefulness, fitness for use and practical limitations of the generated models in everyday tasks. CI-generated BPMN diagrams and associated PNGs were released to ALU, internal demonstrations were held and semi-structured interviews were conducted. Fourteen ALU participants were invited to use the diagrams in their daily support work, after which they participated in 45-minute interviews guided by themes such as current practice, diagram fidelity, usability, coverage, impact on work, integration and traceability, and improvement suggestions. Stakeholders were drawn from both support and development teams and ranged from junior scholarship students to senior technicians. Most had little to no prior exposure to BPMN.

\paragraph*{Findings - ALU (support).}
ALU interviewees reported that the diagrams provided a macro, top-down view that complemented their usual practice of simulating processes in a test environment ("Ashes"). The BPMN models helped them decide whether reversal paths existed and identify the responsible actors without trial-and-error; in one case a diagram revealed a production bug where a student action was present in the model but absent at runtime. For complex cases such as contract proposals, the diagrams allowed support staff to confirm cancellation sequences and actor responsibilities, reducing reliance on ad-hoc simulation. Participants consistently found the BPMN representation easier to grasp than raw JSON, although large processes suffered from overlaps and ambiguous labels. Recommendations included using neutral labels, simplifying gateway names, and providing a legend or cheat-sheet. Interviewees also requested explicit representation of external systems (e.g., SAP, Admissions, Signer), and cross-module user-experience friction emerged as an issue. Requests for change awareness were frequent participants wanted notifications and visual markers when flows changed-and they highlighted the need to improve lane assignments and correct gateway types (e.g., switching from exclusive to parallel splits when multiple processors execute in parallel). For onboarding, diagrams combined with the test environment simulations provided sufficient context; respondents preferred static PNGs over requiring a BPMN editor and requested a one-page legend, a list of fields or a form preview, and test environments for training exercises.

\paragraph*{Findings - development team (DASI).}
Feedback from the development team corroborated and complemented the ALU perspective. Senior developers viewed BPMN or PNG outputs as useful but not essential because they were familiar with the JSON and engine semantics; they valued traceability links for audits but typically used the diagrams opportunistically. Scholarship and junior developers, however, reported high utility: they found request files clear, appreciated the intuitive domain mapping, and benefited from the ability to navigate from BPMN elements to underlying forms and processor code. They identified several form-level issues (e.g., missing flags, inconsistent custom properties, absent justification text boxes) and asked for more visible processor invocation points and better discoverability. The difference between senior and junior usage patterns echoes broader observations about AI-assisted programming: less-experienced developers gain the most from structured guidance and examples, whereas experienced developers use such tools sparingly. Recommendations from the development team included enriching documentation (concise PDFs with annotated screenshots), adding badges or overlays to indicate processor invocation points, providing version diffs, and implementing lightweight validation gates in CI to detect common form-property errors.

\paragraph*{Parallel initiative - SmartIES.}
This prototype module aggregates form and flow information in a unified view; its development was informed by stakeholder sessions where the ontology helped define a common vocabulary and field dictionary. Excel-based field maps served as communication artifacts, helping bridge technical and non-technical stakeholders and exposing process complexity. Early adoption of SmartIES indicated that ontology-backed naming improved discussions and accelerated convergence on new views. While SmartIES lies outside the scope of the transformation pipeline itself, its use underscores the value of ontology-driven abstractions for broader process management.

\paragraph*{Synthesis \& limitations.}
Overall, the qualitative evidence demonstrates that our pipeline meets its intended purpose. Embedding the evaluation in daily operations and working closely with ALU and DASI allowed rapid iteration. The diagrams contributed to faster diagnosis, reduced ad-hoc clarifications and improved escalation quality, and they offered a "zoomed-out" view that facilitated onboarding. Nevertheless, limitations were evident: very large diagrams require manual tidying, static models cannot capture dynamic or time-based behaviors, and some labels and gateways need simplification. The evaluation drew on observational and descriptive techniques and used workflows from a single institution; thus, while the pipeline is designed to be adaptable via new mappings and alignments, generalizability to other domains remains to be tested. Despite these constraints, the validation shows that the ontology-driven transformation provides stakeholders with more understandable and actionable process models than the raw Smart Forms \& Smart Flow definitions. These results underpin subsequent discussions of threats to validity and inform our plans for future work.

\section{Discussion}
\label{sec:discussion}

Transforming the Smart Forms \& Smart Flow definitions into BPMN 2.0 models provides a substantive validation of the proposed ontology‑driven pipeline. Our experiments on 69 real‑world workflows yielded 92 BPMN diagrams with a 94.2~\% success rate and an average translation time of 404~ms per file, demonstrating that the approach is effective, scalable and suitable for continuous integration workflows. The resulting BPMN models improve stakeholder comprehension and maintainability by exposing the implicit control‑flow encoded in JSON and linking tasks, forms and processors back to their sources. Interviews with support staff and developers showed that the diagrams offer a top‑down view that complements the reactive, instance‑based diagnosis used in practice and reduces trial‑and‑error in test environments. For example, the BPMN representation of the contract proposal process (PDEC) enabled a clear understanding of who can act and in which sequence to reverse a request, replacing ad‑hoc exploration and accelerating escalation decisions.

The comparison with Camunda and Oracle implementations underscores that executable BPMN configurations are proprietary across vendors. While BPMN 2.0 is extensible, engine‑specific data mappings and control information are often encoded in extension elements that are not portable. Our ontology‑driven approach localizes this variability in declarative mapping rules and adapter components rather than scattering it across code, enabling the translation of proprietary workflows to standard BPMN while accommodating vendor‑specific details only in clearly defined adapters. This separation improves maintainability and projectability: the Java pipeline remains stable while mappings and ontologies evolve, and new engine profiles can be supported by adding adapters rather than rewriting the transformation logic.

The evaluation also surfaced limitations and boundary conditions. First, the generated diagrams faithfully represent the static structure of Smart Forms \& Smart Flow specifications but cannot capture dynamic or time‑based behaviors that arise only at runtime; 5.81~\% of files could not be translated because they contained transitions or routes computed on the fly. Second, very large processes may exhibit overlapping labels or dense layouts that hinder readability; light manual adjustments are sometimes necessary to tidy the diagrams. Third, the validation focused on structural correctness and organizational fitness‑for‑use; comprehensive code‑level verification of processors and end‑to‑end equivalence between the generated BPMN and runtime execution was outside our scope. Finally, some assumptions remain tied to the Smart Forms \& Smart Flow environment (e.g., queue naming conventions and processor parameters). Generalization is feasible under stated boundary conditions, but adapters must be developed for each new engine, and the ontology and mapping rules must be extended to cover additional constructs and exception patterns. Despite these limitations, the evidence suggests that an ontology‑driven pipeline can systematically bridge proprietary and standard workflow languages, enhancing interoperability and reducing vendor lock‑in.

\section{Conclusion and Future Work}
\label{sec:conclusion}

This work presented a modular, ontology‑driven model‑to‑model transformation pipeline that lifts JSON‑based Smart Forms \& Smart Flow specifications to RDF/OWL using RML, aligns them with a BPMN ontology through reasoning, and generates standards‑compliant BPMN 2.0 diagrams via the Camunda Model API. The pipeline preserves semantic traceability, enabling stakeholders to navigate from BPMN elements back to the original forms and processors, and is integrated into a continuous integration workflow that batch‑converts the entire corpus and publishes artifacts for daily validation. Quantitative evaluation showed high success rates and scalability, while qualitative feedback from practitioners confirmed that the generated diagrams improve understanding, diagnosis and onboarding. Compared with manual modeling or code‑centric converters, the ontology‑driven method reduces the marginal cost of new workflows to near zero after setup and isolates variability in adapters and mappings. These contributions advance the state of the art in ontology‑supported workflow translation and demonstrate the value of design‑science research in bridging proprietary systems with open standards.

\subsection{Future Work}

Several avenues for future work arise from our findings. Extending the ontology and mapping rules to cover additional workflow constructs - such as time‑events, data objects and advanced gateways - will increase coverage and reduce the 5.81~\% failure rate observed. Translating to other modeling standards (e.g., UML Activity Diagrams) and exploring reverse engineering (importing BPMN back to proprietary formats) would further enhance interoperability, though reverse translation demands careful consideration of semantics lost in vendor‑specific extensions. Improving change awareness and readability is another priority: enhanced change‑diff views, explicit representation of external systems via pools and lanes, neutralized labels and simplified gateway phrasing can make diagrams more accessible to non‑technical stakeholders. On the tooling side, priorities include enriching adapters and mappings for broader projectability, improving the layout algorithm to reduce edge crossings and overlaps, packaging diagrams with lightweight viewers (PNG/PDF export and web readers), and evolving continuous‑integration outputs into living documentation (e.g., annotated field maps and screenshots). Finally, integration with platform development processes could be strengthened by adding JSON schema validation and semantic linting to CI pipelines, defining versioned input profiles with migration notes, maintaining per‑process change logs, introducing contract tests for frequently used processors, and instrumenting processor invocation points to correlate runtime traces with BPMN nodes. Pursuing these directions will expand the applicability of the ontology‑driven pipeline, support broader adoption across organizations and vendors, and provide richer feedback loops for both developers and domain experts.

\section{Data and Artifact Availability}

\noindent To facilitate reproducibility and provide additional context for the ontology‑driven workflow translation described in this paper, we supply a \texttt{supplementary\_work.zip} archive. This archive collects all of the diagrams and reference artifacts used throughout our thesis and the evaluation described here. The contents fall into three broad categories:
\begin{itemize}
\item \textbf{Overall architecture and process diagrams.} Files such as \texttt{component\allowbreak\_model.png} (a high‑level component diagram of our translation pipeline), \texttt{functional\_modules\_ist.png} (showing the functional modules deployed at IST), \texttt{sequence\_diagram.png} (an interaction diagram for the transformation steps) and \texttt{source\_target\_mapping.png} (illustrating how source constructs are mapped to BPMN elements) give readers a visual overview of the proposed pipeline.

\item \textbf{Evaluation artifacts.} Under the \texttt{evaluation/} directory you will find performance graphs and supporting materials used to validate our approach. These include time‑complexity plots (\texttt{avg\_time\_vs\_elements.png} and \texttt{avg\allowbreak\_time\_vs\_json\_lines.png}), distributional summaries (\texttt{boxplots.png}), a characterization table summarizing the JSON corpus (\texttt{characterization\allowbreak\_table.png}), an overview of per‑file results (\texttt{json\_files\_results.png}), a matrix report (\texttt{matrix\_report.png}) and two interview guides used in our qualitative assessment (\texttt{interview\_guide\_1.jpg} and \texttt{interview\_guide\_2.jpg}).

\item \textbf{Implementation example.} The \texttt{implementation\_example/} folder contains artifacts illustrating how the ontology is loaded and applied. It provides screenshots from Protégé (\texttt{merged\_ontology\_ex\_protege.png}), images of the semantic lifting of an action node (\texttt{actionnode\_generated\_by\allowbreak\_semantic\_lifting.png}), the corresponding BPMN mapping (\texttt{actionnode\allowbreak\_usertask\_mapping.png} and \texttt{bpmn\_based\_ontology\_usertask.png}), example ontology and ABox files (\texttt{bpmn\_bbo.ttl} and \texttt{smart\_forms\_flow\_ontology\allowbreak\_example.ttl}), pseudocode for the generation routines (\texttt{generateBpmnModel\allowbreak\_pseudo.java} and \texttt{process\_all\_files\_pseudo.java}) and an illustration of our lane‑aware auto‑layout algorithm (\texttt{layout\_alg/lane\_aware\_auto\allowbreak\_layout.png}). There is also a figure (\texttt{comparison\_layout\_kitzmann.png}) comparing our layout with that of Kitzmann et al., as well as images explaining how ontologies and ABoxes are loaded (\texttt{load\_ontologies\_and\_abox.png}).
\end{itemize}

\noindent Additionally, the archive includes a \texttt{demo.pdf} document used in demonstrations and a \texttt{json\_files\_results.png} summarizing the evaluation corpus. All figures referenced in the journal and thesis are contained in this archive to aid replication and further research.

These supplementary materials complement the conclusion by providing the artifacts used to generate the reported results and visual aids that support the narrative of improved comprehension, diagnosis and onboarding.

%
%
%
%
 \bibliographystyle{splncs04}
 \bibliography{mybibliography}

\end{document}